\documentclass[twocolumn]{cinc}
\usepackage{epsfig}
\begin{document}
\bibliographystyle{cinc}

\title{Decomposition of Heartbeat Time Series: Scaling Analysis of the
Sign Sequence}

\author{Y~Ashkenazy$^{1}$, PCh~Ivanov$^{1,2}$,
S~Havlin$^{3}$, C-K~Peng$^{2}$,
Y~Yamamoto$^{4}$, AL~Goldberger$^{2}$, HE~Stanley$^{1}$ \\[1em]
$^1$ Center for Polymer Studies and Department of Physics, Boston
University, Boston, Massachusetts 02215, USA \\[.25ex]
$^2$ Harvard Medical School, Beth Israel Deaconess Medical Center,
Boston, Massachusetts 02215, USA \\[.25ex]
$^3$ Dept. of Physics and Gonda Goldschmied Center, Bar-Ilan University,
Ramat-Gan, Israel \\[.25ex]
$^4$ Tokyo University, Graduate School of Education, Education
Physiol. Lab., Bunkyo Ku, 7-3-1 Hongo, Tokyo 1130033, Japan\\[1em]
}
\maketitle

\begin{abstract}
{ 
The cardiac interbeat (RR) increment time series can be decomposed into 
two sub-sequences: a magnitude series and a sign series. We show that the
sign sequence, a simple binary representation of the original RR series,
retains fundamental scaling properties of the original series, is robust
with respect to outliers, and may provide useful information about
neuroautonomic control mechanisms. 
}
\end{abstract}

\section{Introduction}
Many biological and physical systems exhibit complex dynamics
characterized by long-range correlation properties (scaling laws)
\cite{Shlesinger,Vicsek,Takayasu}. For example,
apparently noisy normal cardiac interbeat ($RR$) interval time series 
obey scaling laws and the RR {\it increment} series shows long-range
anticorrelations \cite{Kobayashi}. Moreover, these scaling exponents
may have diagnostic and prognostic utility \cite{Peng95,Huikuri}. 

\begin{figure}[h!]
\centerline{\epsfig{figure=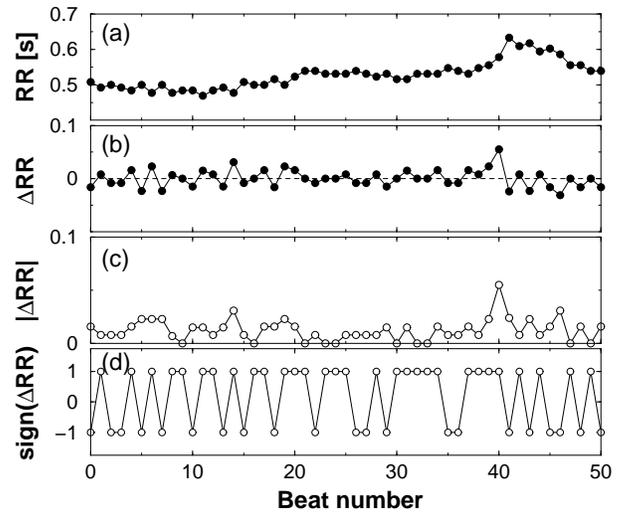,width=7cm,angle=-90}}
\caption{\label{fig2}
An illustration of the magnitude/sign decomposition.
(a) The $RR$ time series (healthy subject) versus beat number for 50
beats.
(b) The increment, $\Delta RR$ of the $RR$ time series shown in (a).
(c) Magnitude of the increments of the
successive heartbeat intervals of the series shown in (a).
Patches of less ``volatile'' increments with small magnitude (beat
number 20-35) are followed by patches of more volatile increments with
large magnitude (beat number 35-50),  
suggesting that there is correlation in the magnitude time series. 
(b) The sign sequence, of the $RR$ series shown in (a). The 
positive sign ($+1$) represents a positive increment, 
while the negative sign ($-1$) represents a negative increment in the $RR$
series of interbeat intervals. The complex
alternation between $+1$ and $-1$ is consistent with the finding 
\protect\cite{ashkenazy} that there is multi-scale anticorrelation in the
sign time series.} 
\end{figure}

\begin{figure}[t]
\centerline{\epsfig{figure=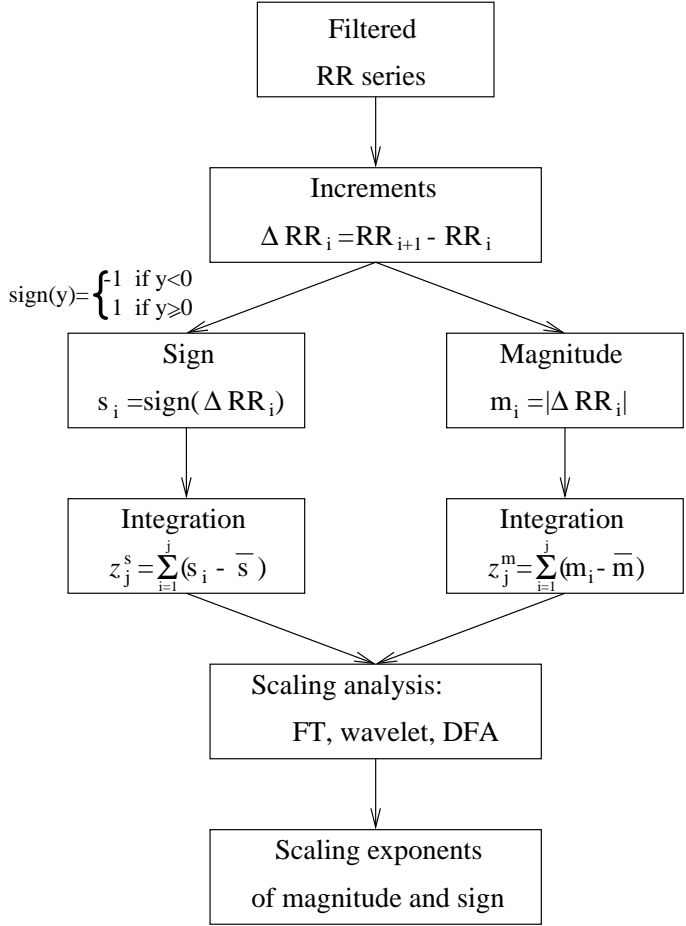,width=9cm,angle=0}}
\caption{\label{fig1}
A schematic representation of the magnitude/sign decomposition of the
$RR$ time series. FT: Fourier transform; DFA: detrended fluctuation
analysis.} 
\end{figure}

In a previous report \cite{ashkenazy} we showed that two components
of the heart 
interbeat interval increment series, namely the magnitude
(``volatility'') and sign, can
be of help in understanding the underlying dynamics which relate to
correlation properties in the original heartbeat time
series (Figs. \ref{fig2} and \ref{fig1}). The heartbeat increment
magnitude series was shown to reveal some
nonlinear aspects of the original heartbeat increment series. On the other
hand, the heartbeat increment sign series mainly reflects the linear
properties of the heartbeat increment series. The correlation properties 
of the sign series may also be of use in separating healthy
patients from heart-failure patients. Surprisingly, the
sign series yields equivalent, and in some cases better
separation of these two groups than the original heartbeat interval time
series. 

Here we focus in further detail on the sign sequence of the $RR$ increment 
series. We show
that sign series is robust in handling complex signals which
include spikes. 
In addition we demonstrate that $\beta$-blockade in healthy young adult
subjects causes an increase in the short range anticorrelated behavior of
the $RR$ increment time series and its component sign series.
Our results, therefore, suggest that the simple sign series can mimic 
the short-range correlation properties of the original $RR$ series and
that it may have advantages when considering noisy data with many outliers.

\section{Method}

The time series of the fluctuations in heartbeat intervals can be
``decomposed'' into two different time series.  This procedure allows for 
separate analysis of the
time sub-series formed by the magnitude and the sign of the increments
in the time intervals between successive heartbeats
(Figs. \ref{fig2} and \ref{fig1}). 

Given a filtered $RR_i$ series we generate the increment
$RR$ series, namely, $\Delta RR_i = RR_{i+1}-RR_i$. Then, the increment
series is decomposed into two sub-series: the magnitude series $m_i=|\Delta
RR_i|$, and sign series $s_i={\rm sign}(\Delta RR_i)$\footnote{The sign
of zero is considered as 1.}. 
Then we integrate the magnitude and sign sub-series.
To avoid artificial trends, prior to the
integration step we subtract from the magnitude and sign series their
averages. Then, we use the detrended fluctuation analysis (DFA) method
\cite{Peng95} to
calculate the root mean square fluctuation function, $F(n)$, where $n$
is the window scale. 

The DFA method excludes nonstationarities which arise because of trends
that are not necessarily related to the cardiac dynamics. The $1^{\rm st}$
order DFA method excludes constant trends that exist in the original
series; the $2^{\rm nd}$ order DFA excludes linear trends that appear in the
data, and higher order of DFA excludes higher order of polynomial
trends. In our 
case, namely, the $RR$ time series, we found that the $2^{\rm nd}$ order
of DFA is adequate\footnote{The $F(n)$ curves are used to approximate
the scaling exponents that characterize the long range correlation of a
given series. The $q^{\rm th}$ order of DFA method can measure
accurately series with scaling exponents, $\alpha$, that are in the
range of $1/2 \le \alpha \le q+1/2$. In the present case the scaling
exponents do not exceed the value of 1.8 and thus the $2^{\rm nd}$ order
of DFA is adequate. Moreover, we have applied higher order DFA to
our data and find similar values for the computed scaling
exponents.}. 

The final step, is to calculate the scaling
exponents of the integrated magnitude and sign series. The scaling exponent,
$\alpha$, is the exponent which quantifies the growth of the root mean
square fluctuations, $F(n) \propto n^\alpha$. 
These steps are schematically summarized in Fig \ref{fig1}.

We note that other methods, such as Fourier transform and wavelet
transform \cite{Ivanov99},
can be applied to calculate the scaling exponents. In these cases
one can avoid the integration step of the magnitude and sign decomposition
(see Fig. \ref{fig1}) and measure the scaling exponents directly from the
magnitude and sign series. However, we use the DFA method because of its
ability to exclude trends and its simplicity.

\section{Examples}
\subsection{Surrogate noise}
In real life examples, like the heartbeat $RR$ time series, outliers 
due to noise, missing beats or extrasystoles that
appear in the recorded signal might alter the scaling exponent
calculation. However, by definition, the sign series does not contain any
outliers, and thus preserves the correct scaling properties despite
the spikes existing in the original signal.

To illustrate this point, we generate complex series with intrinsic
long-range correlations \cite{Makse} (we chose scaling exponent,
$\alpha=0.3$, similar to the long-range scaling exponent that observed
in some $RR$ increment time series). Then, we replace some of the data points with
spikes. We then compare the root mean square fluctuation function $F(n)$ of the
two series, namely, the original long-range correlated series and the
series which contains spikes. We repeat this procedure for the
integrated sign series obtained from the two surrogate series.

\begin{figure}[t!]
\centerline{\epsfig{figure=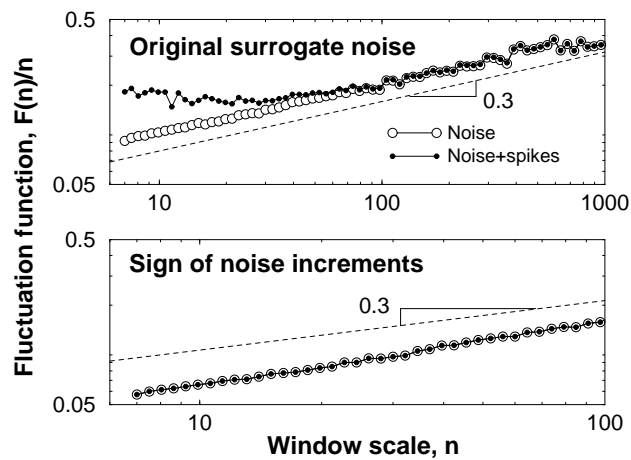,width=6.2cm,angle=-90}}
\caption{\label{fig3}
We generate a long-range correlated series (16384 data points) with
scaling exponent of 
$\alpha=0.3$ for the increment series with and without spikes. Then, 
we calculate the
root mean square fluctuation function $F(n)$ both of the original
surrogate series 
($\circ$) and of the surrogate series with added spikes ($\bullet$). The
upper panel 
shows the results of the original long-range correlated noise; the
noise+spike fluctuation function is very much different than the
original noise fluctuation function. The 
integrated sign series shows identical results for both series. 
The dashed line indicates the slope (scaling exponent) of 0.3.
}
\end{figure}

In Fig. \ref{fig3} we show the results for series of 16384
data points. The spike size is 10 times the standard deviation of the
original surrogate series. We replaced just 5 data points with 
spikes. It is clear from the figure that the short-range correlation
properties of the original surrogate noise are lost due to the appearance
of these spikes, while the integrated sign series preserves the scaling
properties of the original sign series. 

\subsection{$\beta$-Blockade}

\begin{figure*}[tbh]
\centerline{\epsfig{figure=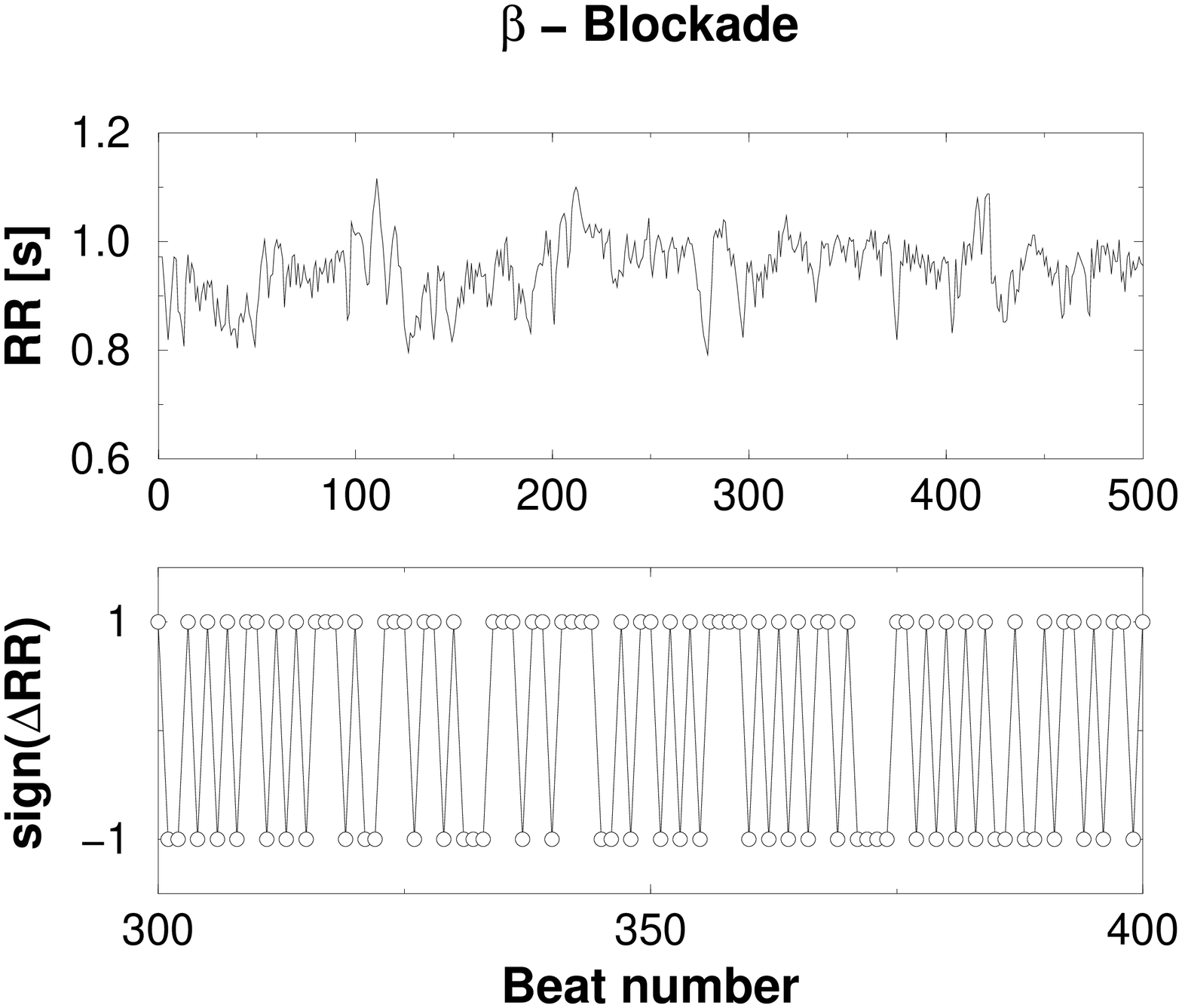,width=7cm,angle=-90}\qquad\epsfig{figure=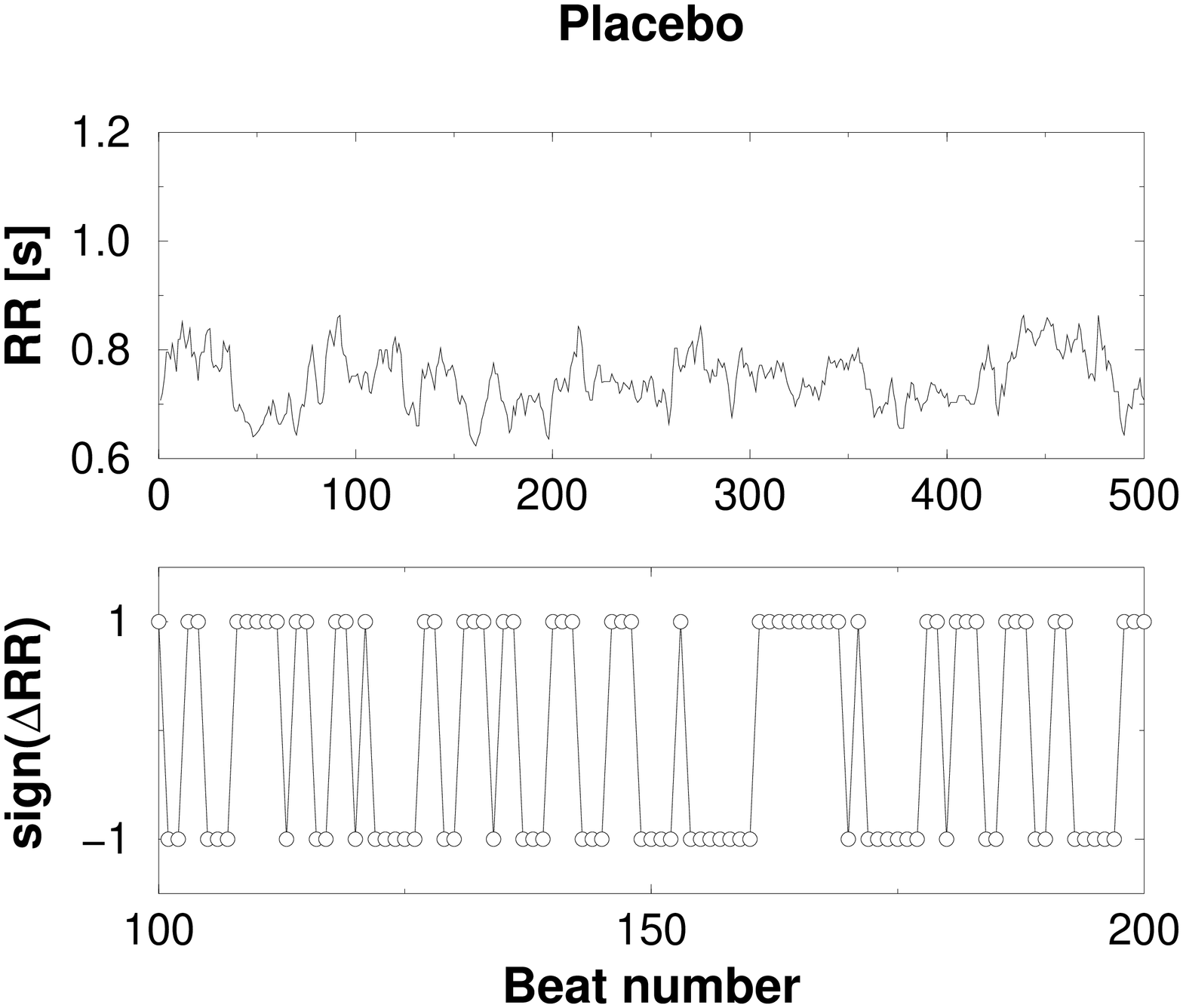,width=7cm,angle=-90}}
\caption{\label{fig4}
Examples of heartbeat interval time series after administration of a
long-acting $\beta$-blocking drug, metoprolol tartrate, 120 mg (left panels) and under normal conditions 
(placebo, right panels) in a young adult healthy subject. The
$\beta$-blockade time series and its component sign 
series show more alternations which suggest stronger
anticorrelated behavior compared with placebo conditions as 
confirmed by quantitative scaling analysis. See Fig. \protect\ref{fig5}.
}
\end{figure*}

Healthy subjects under the influence of
$\beta$-blockade may exhibit increased  high frequency $RR$ behavior
\cite{Yamamoto}. This prominent
high frequency behavior is consistent with the strong
alternations and large variability observed in the data as compared to
placebo control (Fig. \ref{fig4}). 
The alternations of the heart interbeat
increments can readily visualized, in the sign series. These
alternations suggest stronger anticorrelated behavior (for both the $RR$
series and the sign series) under the influence of the $\beta$-blocking
drug. 

\begin{figure*}[tbh]
\centerline{\epsfig{figure=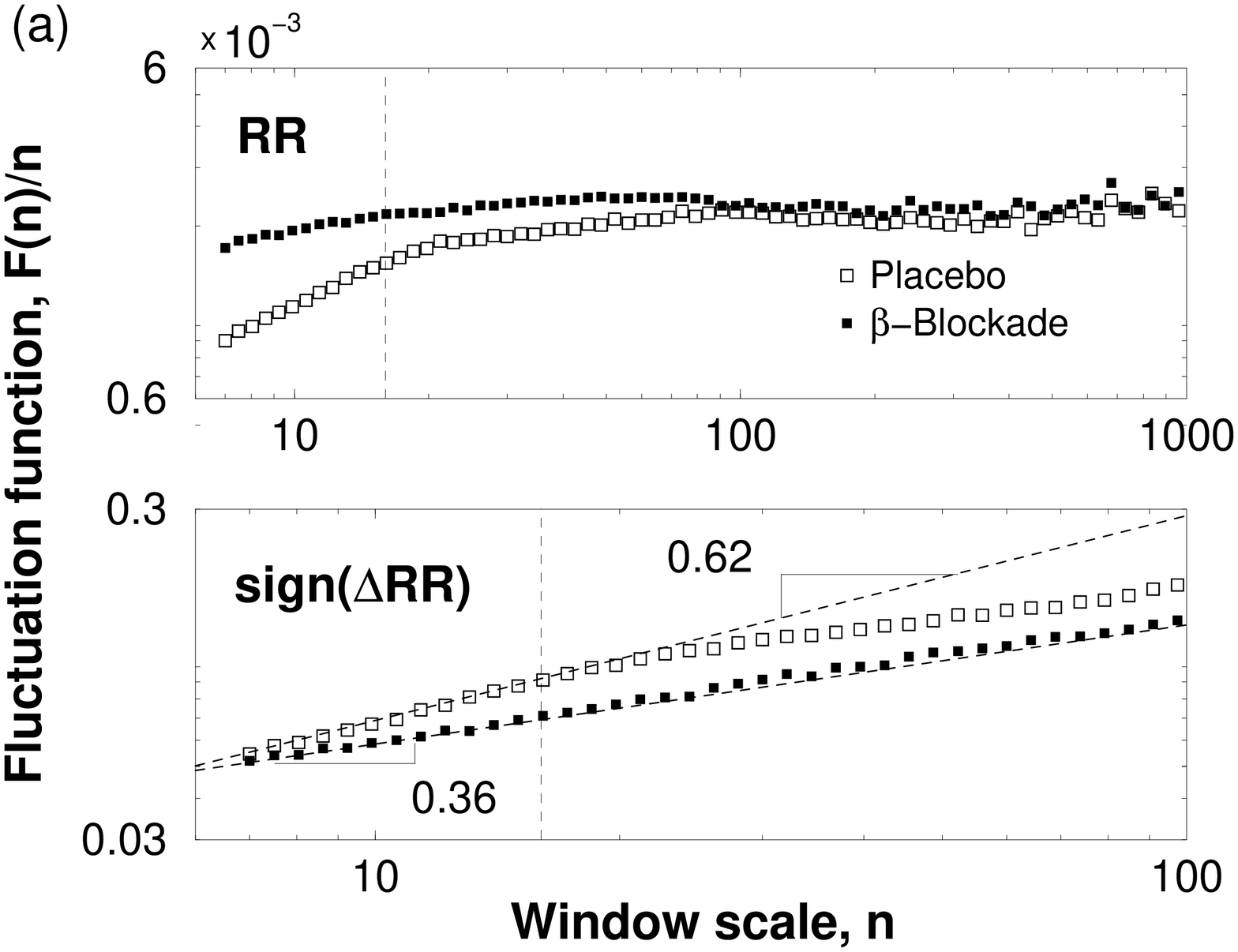,width=7cm,angle=-90}\qquad\epsfig{figure=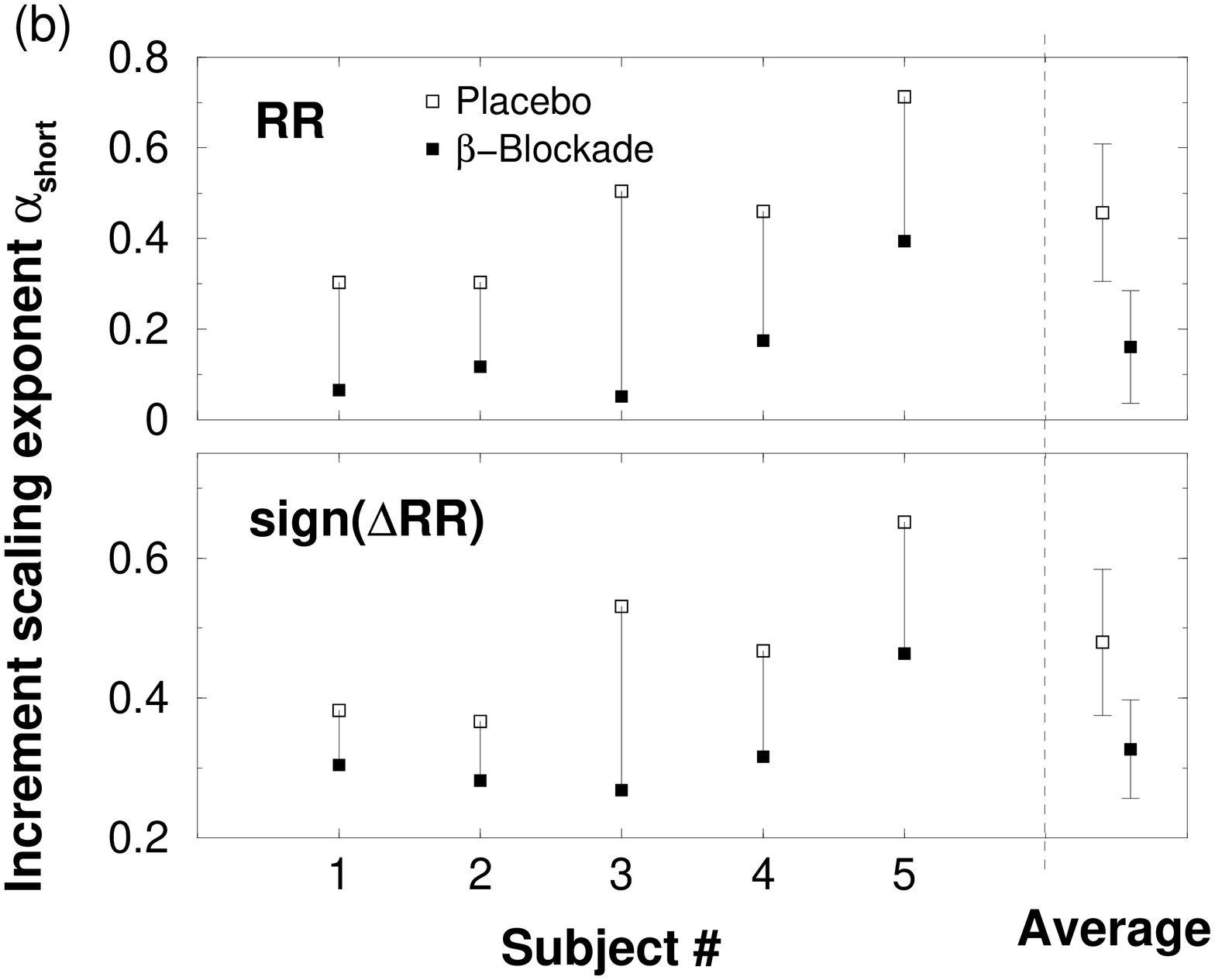,width=7cm,angle=-90}}
\caption{\label{fig5}
(a) Effect of $\beta$-blockade in healthy young subjects.
The scaling of the original $RR$ time series changes at short time scales
and instead of an apparent crossover at time scale $n \approx 16$ beats
(indicated by the dashed vertical line), we have a single
scaling exponent. 
For the sign series we observe the same changes as in the original data. 
(b) A summary of 14 records, 30000 data points each, from 5
subjects. The subjects were recorded 
either with placebo or $\beta$-blockade therapy (subjects 2 and 5 have 2  
records each both for $\beta$-blockade and placebo; the average value is
given.) The figure  
shows a systematic decrease of $\alpha_{\rm short}$, the short range
scaling exponent ($7 < n < 16$ beats) both for the
original $RR$ series as well as for the sign series. This finding suggests
that the short-range scaling exponent 
relates to the interaction between the sympathetic and the parasympathetic
components of the neuroautonomic nervous system. The error bars indicate
the mean $\pm$ 1 standard deviation. 
}
\vspace*{-0.3truecm}
\end{figure*}

Our correlation analysis indeed shows stronger anti-correlated behavior
after $\beta$-blockade for $RR$ series as well as for the sign series
(Fig. \ref{fig5}). These results suggest that interaction between the
sympathetic and the parasympathetic systems is reflected by the sign of
the heartbeat increments.

\section{Summary}

We study one of the most basic representations of cardiac interbeat increment
dynamics --- the sign series. We show that it 
is robust to outliers which may appear in the heartbeat time
series. Moreover, we 
show that the sign series reflects changes in short-range neuroautonomic
effects. The sign series may add additional information 
and may aid in cardiac diagnosis and prognosis in certain conditions. 

\section*{Acknowledgments}
{\small Partial support was provided by the NIH/National Center for Research
Resources (P41 RR13622) the G. Harold and Leila Y. Mathers Charitable
Foundation, the Fetzer Foundation, and the Israel-USA Binational Science
Foundation.} 

\bibliography{psd} 

\begin{correspondence}
\vspace*{-0.1truecm}
Y. Ashkenazy\\
Center for Polymer Studies and Department of Physics, Boston
University, Boston, Massachusetts 02215, USA \\
ashkenaz@argento.bu.edu
\end{correspondence}
\end{document}